# A Trapped Field of 17.6 T in Melt-Processed, Bulk Gd-Ba-Cu-O Reinforced with Shrink-Fit Steel

Durrell, J.H.[1], Dennis A. R.[1], Jaroszynski J.[2], Ainslie, M. D.[1], Palmer, K. G. B.[1], Shi Y-H[1], Campbell, A. M.[1], Hull, J.[3], Strasik, M.[3], Hellstrom, E.[2], Cardwell, D. A.[1]

1. Department of Engineering, University of Cambridge, Trumpington Street, Cambridge, UK.
2. Applied Superconductivity Center, National High Magnetic Field Laboratory, Florida State University, 2031 East Paul Dirac Drive, Tallahassee, Florida 32310, USA.
3. Boeing Co., Seattle, USA.

## Abstract

The ability of large grain, $REBa_2Cu_3O_{7-\delta}$ [(RE)BCO; RE = rare earth] bulk superconductors to trap magnetic field is determined by their critical current. With high trapped fields, however, bulk samples are subject to a relatively large Lorentz force, and their performance is limited primarily by their tensile strength. Consequently, sample reinforcement is the key to performance improvement in these technologically important materials. In this work, we report a trapped field of 17.6 T, the largest reported to date, in a stack of two, silver-doped GdBCO superconducting bulk samples, each of diameter 25 mm, fabricated by top-seeded melt growth (TSMG) and reinforced with shrink-fit stainless steel. This sample preparation technique has the advantage of being relatively straightforward and inexpensive to implement and offers the prospect of easy access to portable, high magnetic fields without any requirement for a sustaining current source.

## Introduction

It has long been known that, in addition to fabricating solenoids from wire or tape, type-II superconducting materials can be used to trap magnetic fields when fabricated in the form of well-connected bulks [1, 2]. Top-seeded melt growth (TSMG) has emerged over the past 25 years as a practical route for fabricating large, single grains of the rare earth (RE) cuprate family of high-temperature superconductors (HTS) of composition $(RE)Ba_2Cu_3O_{7-\delta}$ [(RE)BCO]. As a result, these materials have significant potential for application, effectively, as high-field permanent magnets [3]. The performance of these magnets at 77 K is limited by the critical current carrying capacity of the bulk superconductor. Nevertheless, fields of up to 2 T have been achieved in 20 mm diameter superconducting bulk samples [4] and up to 3 T in samples of diameter 65 mm [5] at 77 K.

The critical current density ($J_c$) of HTS is enhanced at temperatures lower than 77 K, and significantly larger magnetic fields can be trapped. Notably, Tomita and Murakami reported a trapped field of 17.24 T at 29 K in a $YBa_2Cu_3O_{7-\delta}$ (YBCO) sample of 26 mm diameter impregnated with Wood's metal and resin and reinforced with carbon fibre [6]. Fuchs et al. also reported a trapped field of 16 T at 24 K in a Zr doped and Ag impregnated YBCO sample of 25mm diameter placed inside a reinforcing stainless-steel tube [7]. The prospect of generating portable high fields that are available outside the bore of a superconducting solenoid is now a distinct possibility given that considerable progress is being made in developing practical systems for magnetising and refrigerating such samples to operating temperatures well below 77 K [8]. Additionally, these superconducting bulk permanent magnets operate in what is effectively "persistent mode", which cannot yet be achieved in solenoids fabricated from YBCO coated conductors, due to the difficulty in making superconducting joints.

There are significant challenges, beyond merely producing a material that exhibits a sufficiently large critical current density, to trap very high fields in (RE)BCO bulk superconductors. Firstly, the motion of flux during the ramp-down of any applied, external magnetic field generates a significant amount of heat, and this can lead to the formation of thermal instabilities and consequent catastrophic quenching of the bulk superconductor [6]. Secondly, the forces exerted on the sample during magnetisation due to the Lorentz force are large and can lead to mechanical failure with unreinforced samples typically failing at ~7-8 T [9, 10]. The forces in a bulk sample during the ramp-down of the magnetising field are tensile in nature [11], and the resulting tensile stresses are not resisted well by the brittle ceramic-like nature of the material. Moreover, as it is the resistance to fracture that determines the tensile performance and, given that the TSMG process generates many voids and cracks in the sample microstructure, a wide variation in field-trapping performance is to be expected from sample to sample [12].

The aim of the work presented here was to develop a superconducting bulk system that could exhibit peak trapped fields as large, or larger, than those currently reported in the literature using a conventional sample processing technique and reinforcement techniques that are relatively simple to implement.

## Methods

### Reinforcement Analysis

We have developed Ag-containing GdBCO samples that exhibit good field-trapping performance at 77 K [13] and contain a more homogenous $J_c$ distribution than YBCO [14]. Unlike the Ag-GdBCO samples fabricated by other groups, we do not observe a significant deleterious effect on $J_c$ for an Ag content of up to 15 wt% $AgO_2$. This is a consequence of the large number of voids and cracks generated during the TSMG process, which can accommodate the excess Ag without affecting the properties of the continuous superconducting matrix. As a result, it was not necessary to enhance the pinning performance of the materials studied here by introducing an extra pinning phase to the Gd-123 microstructure.

Previous investigators have used stainless-steel rings to reinforce bulk superconductors, relying on the pressure that occurs at the superconductor/stainless steel interface due to differential thermal contraction on cooling from room temperature to the desired measurement temperature. Here, we first determined the magnitude of the interface pressure that would result on the samples using a standard method [15] for analysing the shrink-fitting process and the published values for the mechanical properties of (RE)BCO [16]. It was found that this would equate to a relatively modest 55 MPa for a reinforcing ring of 24 mm diameter and 3 mm thickness cooled to 28 K, which is the same order of magnitude as the tensile strength anticipated in a GdBCO superconducting bulk containing Ag [10]. The required tensile strength of a material for a particular maximum field can be estimated [6] as 0.282 $B^2$ MPa. The required tensile strength at a field of 18 T is therefore ~ 91 MPa, which would result in the mechanical failure of most samples since the internal compressive stress will be less than the interface pressure. The samples were machined precisely to a diameter of 24.15 mm and were reinforced by a stainless steel ring of internal diameter 24 mm to increase the pre-stress. The rings were heated to > 300 ºC to enable them to fit onto the superconductor. This configuration was calculated to provide a pre-stress of ~ 250 MPa, which is a significant improvement on the stress achieved from simple steel banding. It should be noted that, while this compressive stress is added uniformly, the sample interior may not experience a homogeneous stress distribution due to the presence of voids in the sample, which may lead to specific regions of much lower compressive stress.

## Sample preparation and measurements

The cylindrical bulk samples of dimensions ~ 25 mm diameter and ~ 13 mm height were grown using the top-seeded melt growth process described elsewhere [13] and then machined to a diameter of 24.15 mm. The samples exhibited trapped fields of ~ 0.9 T at 77 K. A 304 stainless-steel ring was then heated to a temperature > 300 °C and shrink-fitted to each sample.

The samples were combined into stacks of two, joined by a thermally conductive epoxy resin, as shown in Fig. 1, for measurement purposes. A linear array of five Lakeshore HGT-2101 Hall sensors was placed at the centre of each stack at positions –8 mm, –4 mm, 0, 4 mm and 8 mm from the centre of the sample (i.e. between the two halves). These sensors were supplied un-calibrated and exhibited some non-linearity, which was characterised at 100 K. The output voltage of the Hall sensor was measured at the temperature at which the magnetising field was applied. The temperature of the two-sample stack was measured using a Cernox sensor and the sample temperature was stabilised using a wire-wound heater wrapped around the stack.

The Hall sensors were driven with a 100 Hz, 10 µA peak sine wave using a Keithley 6221 current source. The Hall voltage was measured using a lock-in amplifier for each Hall probe. The samples were magnetised in the bore of the SCM-2 18 T superconducting magnet at the National High Magnetic Field Laboratory, University of Florida. The desired magnetising field was applied with the temperature of the stacks maintained at 100 K, after which the samples were cooled to the measurement temperature. Finally, the external field was ramped down at a rate of 0.015 T/min.

## Results and Discussion

Two similar stacks were magnetised initially by field cooling from 16 T at 28 K. One stack suffered some cracking and trapped a maximum field of 10 T, whereas the second two-sample stack trapped a field of 15.4 T.

A third sample was then magnetised with an applied field of 17.8 T at 26 K. The fields recorded during the magnetising process of this stack are shown in Fig. 2. A trapped peak field of 17.6 T was achieved at the end of the magnetising process, which represents the largest field trapped in a bulk superconductor at any temperature reported in the literature to date.

To monitor any reduction in trapped field due to flux creep, the sample was maintained at 26 K for a period of 160 minutes, as shown in Fig. 3. Flux creep is observed, the rate of which decreases with time, as expected [17]. The flux creep behaviour over the relatively short time period measured is initially logarithmic but then drops more quickly.

The sample was then warmed slowly at a rate of 0.5 K/min and the variation of the trapped field with temperature recorded. Fig. 4(a) shows the field at the centre of the sample and Fig 4(b) the evolution of the trapped field profile with increasing temperature. This data set suggests that the critical current performance of the samples employed would potentially support the maximum trapped field of 17.6 T at temperatures of up to 32 K. It is immediately apparent from the variation of $J_c$ with distance that the field trapped in this sample is not limited by the current carrying capability of the superconductor, as the profile is flatter than would be expected for full penetration. Furthermore, the sample was still capable of trapping almost 10 T at 50 K, a temperature which, significantly, is achievable using a single stage cryocooler. Fig 4(b) shows that the trapped field profile changes to one corresponding to a sample limited by critical current as the temperature of the two-sample stack increases. It is apparent in Fig 4(b) that the centre hall probe was not aligned fully with the centre of the field distribution. This may have been due to slight asymmetry of the superconducting properties of the sample, or slight to misalignment of the two components of the stack.

The sample was then re-magnetised in a separate, but otherwise identical process, using an 18 T field, but failed due to cracking as the external field was reduced. Similarly a further, fourth sample stack failed under a magnetising field of 18 T, suggesting that the maximum trapped field observed in this study corresponds approximately to the limit of mechanical performance for the shrink-fit reinforcement method employed to reinforce the samples.

A spread in performance of four sample stacks was observed in this study for similar magnetising processes. This is not surprising, and indeed is characteristic of a brittle material, which fails typically by fast fracture. Although the Weibull modulus for GdBCO fabricated by TSMG is not available in the literature, the TSMG process is one that is known to generate a large number of cracks of different lengths in large, single grain samples. It is expected, therefore, that wide variation in sample performance will be observed between apparently similar samples from the same growth batch. Statistically, this also means that samples with a larger volume will exhibit poorer performance [18].

The question remains why, given the large predicted interface pressure, samples still failed at stresses lower than the applied pre-stress. The answer to this is likely to be two-fold; the difficulty in providing a well-defined and smooth (RE)BCO surface to mate with the reinforcing ring and the fact that the internal pre-stress provided is somewhat smaller than the interface pressure. This could be addressed by using a larger shrink-fit temperature, although this would risk damaging the samples. The choice of an alternative reinforcement material to 304 Stainless Steel is limited given that this material exhibits an unrivalled combination of stiffness and yield strength, while still possessing a significant coefficient of thermal expansion. Another route would be to fabricate GdBCO with enhanced fracture toughness, perhaps via the use of techniques such as fibre reinforcement to bridge cracks as they form during melt processing.

## Conclusion

The best-performing two-sample stack of bulk GdBCO containing Ag trapped a peak magnetic field in excess of 17.6 T from a 17.8 T magnetising field at 26 K. This is the highest field trapped in a bulk superconductor reported to date at any temperature.

The applicability of the relatively simple technique of applying a shrink-fit steel reinforcement to the samples has been demonstrated clearly. This is a well-understood and easy to implement technique that is applied regularly to ceramic materials [19]. Apart from the addition of Ag and consequent minor adjustment of the thermal process during growth, the GdBCO samples were prepared in air by a conventional cold-seeding top seeded melt process. The technique described here, therefore, provides a practical route to generating very high fields to be trapped in single grain, (RE)BCO superconducting bulk samples. The ceramic nature of these samples leads to a wide variation in fracture toughness and the use of compressive pre-stress is an effective route to enhanced field trapping performance.

## Acknowledgements

This work was supported by the Boeing Company and by the Engineering and Physical Sciences Research Council [grant number EP/K02910X/1]. Part of this work was performed at the National High Magnetic Field Laboratory, which is supported by National Science Foundation Cooperative Agreement No. DMR-1157490, the State of Florida, and the U.S. Department of Energy.

Figures

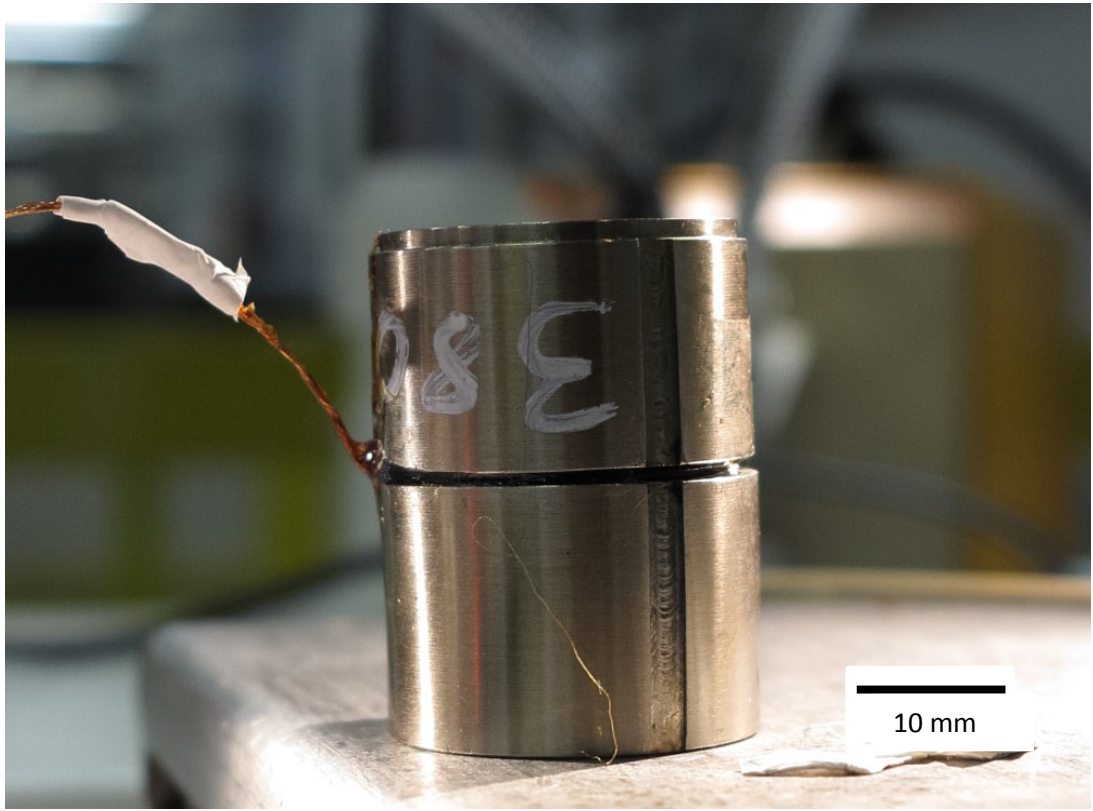

Figure 1. The assembled stack of two GdBCO bulk samples, each of diameter 24.15 mm and height 15 mm. The samples were reinforced with a ring of 3 mm thickness fabricated from 304 Stainless Steel. The recess visible in the ring served to accommodate the measurement wiring. The two bulk samples were then assembled into a stack with an array of 5 Hall probes at their interface using STYCAST thermally conductive epoxy resin. The Hall probes were arranged evenly in a line across the sample at -8 mm, -4 mm, 0, 4 mm and 8 mm from the sample centre.

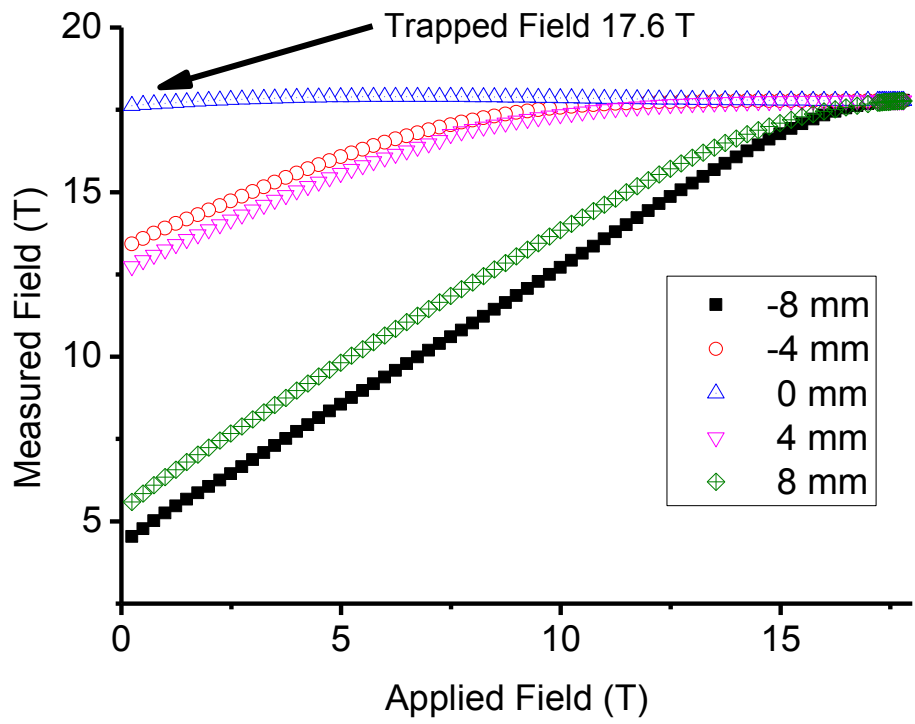

Figure 2. The field measured at the interface between the two samples in the stack by an array of Hall probes (distances indicated are from the centre of the sample) as the magnetising field was ramped down.

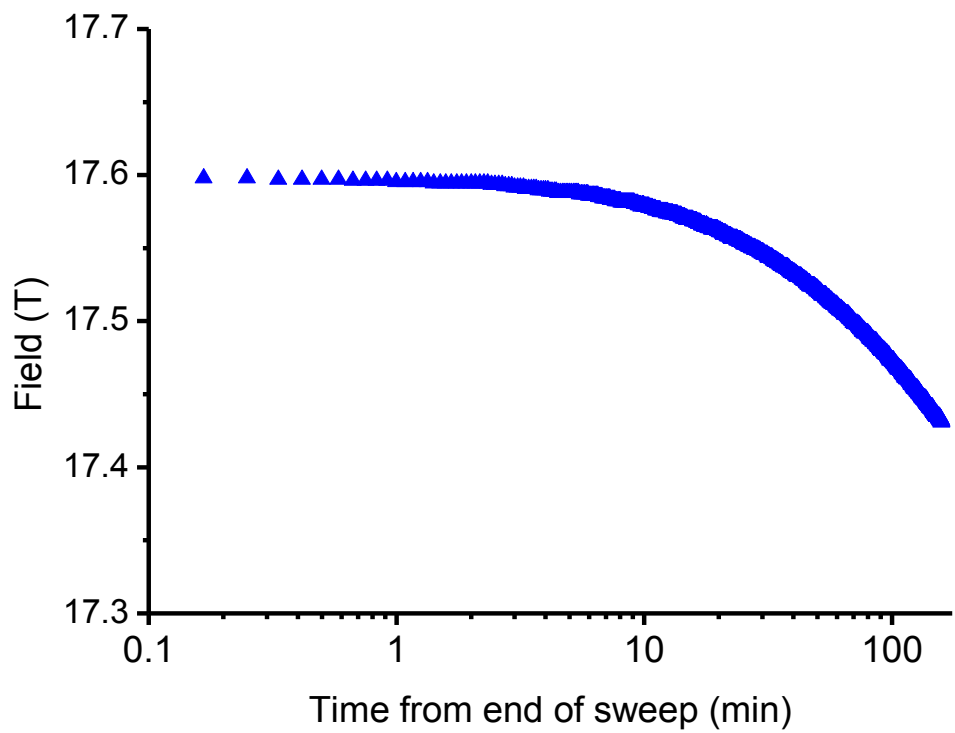

Figure 3. The reduction in trapped field due to flux creep at the end of the magnetising sweep with the sample temperature maintained at 26 K.

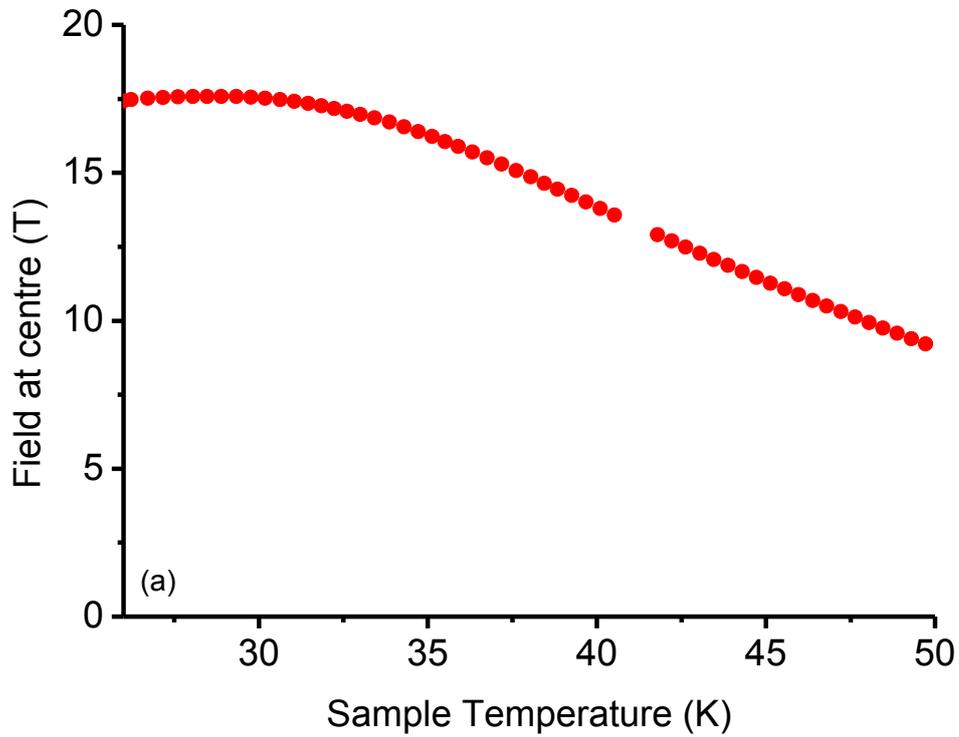

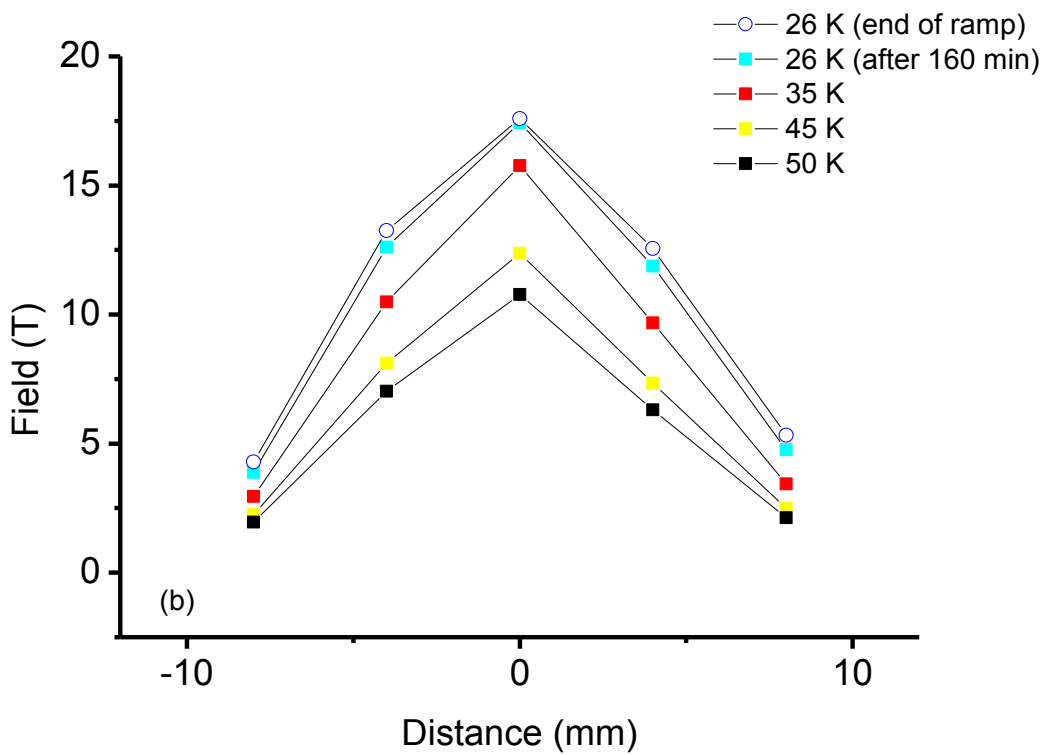

Figure 4. (a) The trapped field measured by the centre Hall sensor as the sample was warmed slowly. (b) The field profile across the sample measured by the five Hall probes 160 minutes after magnetising and at various temperatures during warming of the sample.